\documentclass[reprint, superscriptaddress, nofootinbib, amsmath,amssymb, aps, prc, a4paper]{revtex4-2}

\usepackage{graphicx}
\usepackage{dcolumn}
\usepackage{bm}
\usepackage{nameref}
\usepackage[colorlinks,linkcolor=blue,citecolor=blue,filecolor=black,urlcolor=blue]{hyperref}
\usepackage{orcidlink}
\usepackage{float}
\usepackage{bm}
\usepackage{xcolor}

\newcommand{\snn}{\sqrt{s_{\rm NN}}}
\newcommand{\kstar}{\mathrm{K}^{*0}}
\newcommand{\Tch}{\rm T_{ch}}
\newcommand{\Tann}{\rm T_{ann}^{frz}}
\newcommand{\Tkin}{\rm T_{kin}}
\newcommand{\Nch}{\rm \langle dN_{\text{ch}}/d\eta \rangle^{1/3}}
\newcommand{\pbar}{\rm \overline{p}}
\newcommand{\pt}{p_{\rm T}}
\newcommand{\pann}{\rm (\overline{p}/p)_{T_{ann}}/(\overline{p}/p)_{T_{had}}}

\begin{document}


\title{Exploring the properties of the Hadronic Phase in Heavy-Ion Collisions at RHIC Energies via Partial Chemical Equilibrium}

\author{Rishabh Sharma~\orcidlink{0000-0003-0698-7363}}
\email{rishabhsharma@students.iisertirupati.ac.in}
\affiliation{Department of Physics, Indian Institute of Science Education and
Research (IISER) Tirupati, Tirupati 517619, Andhra Pradesh, India}

\author{Chitrasen Jena~\orcidlink{0000-0003-2270-3324}}
\email{cjena@iisertirupati.ac.in}
\affiliation{Department of Physics, Indian Institute of Science Education and
Research (IISER) Tirupati, Tirupati 517619, Andhra Pradesh, India}

\author{Volodymyr Vovchenko~\orcidlink{0000-0002-2189-4766}}
\email{vvovchenko@uh.edu}
\affiliation{Department of Physics, University of Houston, 3507 Cullen Blvd, Houston, Texas 77204, USA}

\begin{abstract}
The hadronic phase in heavy-ion collisions plays a crucial role in shaping the final-state hadron abundances. In this work, we study Au+Au collisions at $\snn = 7.7$--$200$\,GeV using the Hadron Resonance Gas model in Partial Chemical Equilibrium (HRG-PCE). By fitting the yields of stable hadrons and short-lived resonances such as K$^*(892)^0$, we extract both chemical and kinetic freeze-out temperatures as functions of center-of-mass energy and centrality. The analysis, performed using the \texttt{Thermal-FIST} package, avoids assumptions about radial flow profile or freeze-out hypersurfaces. Furthermore, we estimate the baryon annihilation freeze-out temperature from the experimentally measured $\pbar/$p ratio, using the HRG-PCE framework extended to include $B\overline{B} \leftrightarrow n\pi$ reactions. The inferred annihilation freeze-out temperature lies between the chemical and kinetic freeze-out temperatures, suggesting that baryon annihilation remains active in the early hadronic phase but ceases prior to kinetic freeze-out. These results provide a consistent picture of the sequential decoupling of hadronic processes and demonstrate that inelastic hadronic interactions significantly influence the chemical composition of the system between chemical and kinetic freeze-outs at RHIC energies.

\end{abstract}

\maketitle


\section{Introduction}
\label{sec:intro}

Ultra-relativistic heavy-ion collisions at facilities such as the Relativistic Heavy Ion Collider (RHIC) provide a unique opportunity to study strongly interacting matter under extreme conditions. At sufficiently high collision energies, they are believed to create a deconfined quark–gluon plasma (QGP), which subsequently cools and hadronizes into a ``fireball" of hadrons and resonances.~\cite{STAR:2005gfr,Aoki:2006we,Ejiri:2008xt,ALICE:2022wpn}. As the hadronic fireball expands and cools, it undergoes two distinct freeze-out stages: chemical freeze-out, when inelastic interactions stop and particle yields are fixed, and kinetic freeze-out, when elastic interactions stop and the final momentum distributions are established~\cite{Chatterjee:2015fua}.

Traditionally, experimentally measured hadron yields are simultaneously fit using the Hadron Resonance Gas (HRG) model to determine chemical freeze-out parameters: chemical freeze-out temperature ($\Tch$), baryon chemical potential ($\mu_B$), and radius of the hadronic fireball (R)~\cite{Cleymans:1998fq,Becattini:2009sc,Andronic:2017pug}. On the other hand, the kinetic freeze-out is studied using the blast-wave model, which assumes that the hadronic fireball can be described as an expanding, locally thermalized source with a radial flow profile~\cite{Schnedermann:1993ws}. In this approach, transverse momentum ($\pt$) spectra of various hadrons are simultaneously fit to extract the kinetic freeze-out temperature ($\Tkin$) and radial flow velocity $\langle \beta_{\rm T} \rangle$. However, an anti-correlation between $\Tkin$ and $\langle \beta_{\rm T} \rangle$ emerges, reflecting the interplay between thermal motion and collective expansion in shaping the $\pt$ spectra~\cite{ALICE:2013mez,STAR:2017sal,Tariq:2024hfc}. This anti-correlation could make it challenging to reliably extract the kinetic freeze-out temperature, and the results are dependent on the assumptions about the flow profile.

Between these two freeze-out stages, the hadronic phase may remain chemically active for certain hadrons. In particular, baryon-antibaryon annihilation processes are particularly relevant due to their large cross sections ($>$ 60 mb) and have the potential to significantly affect the final-state baryon yields, including the antiproton-to-proton ratio~\cite{Bass:2000ib,Becattini:2012xb,Steinheimer:2017vju,Vovchenko:2022xil,Savchuk:2021aog}. This effect is expected to be most prominent in central collisions, where the hadronic phase is longer-lived and relatively denser. 

Resonances, such as the $\rho^0$ (lifetime $\sim$1.3 fm/$c$) and $\kstar$ (lifetime $\sim$4.2 fm/$c$), play an important role in freeze-out studies. Due to their short lifetimes, these resonances decay in the hadronic phase, where hadronic re-scattering and regeneration processes occur between chemical and kinetic freeze-out. As a result, their yields and momentum spectra can carry imprints of both freeze-out stages~\cite{Das:2026qnu,STAR:2004bgh,ALICE:2014jbq,STAR:2022sir,ALICE:2018qdv,ALICE:2016sak,ALICE:2023ifn}. These hadronic interactions modify the experimentally observed resonance yields, thereby providing a sensitive probe of the hadronic phase lifetime and its thermodynamic evolution~\cite{Brown:1991kk,Singha:2015fia,Chabane:2024crn,Neidig:2025xgr,Sahoo:2023dkv}.

Experimental measurements have shown that the yields of short-lived resonances are suppressed relative to the HRG model predictions owing to the re-scattering of resonance decay products in the hadronic phase~\cite{ALICE:2019xyr,STAR:2022sir}. To incorporate this effect, the HRG model has been extended to include Partial Chemical Equilibrium (PCE)~\cite{Bebie:1991ij,Motornenko:2019jha,Vovchenko:2019aoz}. In the HRG-PCE framework, the yields of short-lived resonances evolve according to the law of mass action during the hadronic phase, whereas the yields of hadrons stable under strong interactions (hereafter referred to as stable hadrons) remain fixed at their chemical freeze-out values. This results in a state of PCE, wherein short-lived resonances continue to decay and regenerate until kinetic freeze-out. As proposed in Ref.~\cite{Motornenko:2019jha}, the measured yields of such resonances can serve as a probe for estimating $\Tkin$. The underlying assumption that reactions that maintain kinetic equilibrium are dominated by resonance decay and regeneration—a picture that is generally supported by hadronic afterburner simulations~\cite{Steinheimer:2017vju,Oliinychenko:2021enj}.

Using the \texttt{Thermal-FIST} package\footnote{https://github.com/vlvovch/Thermal-FIST} ~\cite{Vovchenko:2019pjl}, which includes the HRG-PCE framework, simultaneous fits to the measured hadron and resonance yields were performed, enabling the extraction of both $\Tch$ and $\Tkin$, along with $\mu_B$ and R without relying on the assumption of flow profile, as is typical in blast-wave analyses, and avoiding the degeneracy between temperature and flow velocity parameters.

Recent developments have further extended HRG-PCE to include baryon-antibaryon annihilation and regeneration processes~\cite{Vovchenko:2022xil}. This makes it possible to determine the annihilation freeze-out temperature ($\Tann$), corresponding to the stage where baryon-antibaryon annihilation reactions effectively cease. In particular, the suppression of the antiproton-to-proton ($\pbar$/p) ratio via reactions such as $B\overline{B} \leftrightarrow n\pi$~\cite{Cassing:1999es,Lin:2004en} in central collisions compared to the peripheral collisions can be used to extract $\Tann$ by contrasting experimental data with HRG-PCE predictions that incorporate annihilation effects in the hadronic phase.

The primary objective of this study is to determine the chemical, baryon-antibaryon annihilation, and kinetic freeze-out conditions by analyzing the measured yields of various hadrons and resonances in Au+Au collisions at $\snn$ = 7.7--200~GeV, using the HRG-PCE model as implemented in the \texttt{Thermal-FIST}. The paper is organized as follows: Sec.~\ref{sec:pce} describes the implementation of PCE in the \texttt{Thermal-FIST}. In Sec.~\ref{sec:results}, we present our results and summarize our findings in Sec.~\ref{sec:summary}.


\section{Partial Chemical Equilibrium}
\label{sec:pce}

The hadronic phase in relativistic heavy-ion collisions can be modeled as a thermal medium governed by a rich interplay of chemical and kinetic processes. The HRG-PCE model provides a thermodynamically consistent framework to describe the late-stage evolution of this phase, where inelastic reactions have ceased but elastic and pseudo-elastic scatterings continue to maintain kinetic equilibrium. In this study, we use the HRG-PCE formulation implemented in the \texttt{Thermal-FIST v1.5.2} to study two key aspects of the hadronic stage: (i) the extraction of chemical and kinetic freeze-out properties via simultaneous fits to stable hadrons and short-lived resonances, and (ii) the role of residual baryon-antibaryon annihilation in shaping the final state (anti-)baryon yields. Both analyses employ the same underlying model assumptions, enabling a unified interpretation of the hadronic phase in high-energy collisions.

\subsection{Kinetic Freeze-out from Resonance Suppression}
\label{subsec:PCE}

After chemical freeze-out at temperature $\Tch$, the system continues to evolve under elastic and pseudo-elastic interactions. In particular, processes that proceed through intermediate short-lived resonances such as $\rm \pi K \to K^* \to \pi K$ and $\rm p K^{-} \to \Lambda^* \to p K^{-}$ remain active. In PCE, these processes are assumed to obey the law of mass action, leading to a gradual suppression of resonance yields as the system cools toward kinetic freeze out~\cite{Vovchenko:2019aoz}. 

To account for this, the model introduces effective chemical potentials for all hadrons~\cite{Motornenko:2019jha}:
\begin{equation}
    \tilde{\mu}_j = \sum_i \langle n_i \rangle_j \mu_i,
\end{equation}
where $i$ runs over all stable and long-lived hadrons whose yields are fixed at T = $\Tch$, $\mu_i$ are their chemical potentials, and $\langle n_i \rangle_j$ denotes the average number of stable hadrons $i$ produced in the decay of resonance $j$. The evolution of the system in PCE is governed by the conservation of stable hadron yields and the entropy of the system:
\begin{align}
    \sum_j \langle n_i \rangle_j \, n_j(\rm T,\tilde{\mu}_j) V &= N_i^{\text{tot}}(\Tch), \\
    \sum_j s_j(\rm T,\tilde{\mu}_j) V &= S(\Tch),
\end{align}
where $n_j$ and $s_j$ are the number and entropy densities of species $j$, and V is the system volume. Solving these equations determines the temperature-dependent $\tilde{\mu}_j$ and V, which are then used to calculate the thermal densities of all hadrons and resonances in the HRG-PCE framework~\cite{Vovchenko:2019aoz}.

As the yields of short-lived resonances are not conserved in PCE, their suppression below $\Tch$ can be used to estimate the kinetic freeze-out temperature $\Tkin$. To this end, we perform simultaneous fits to the mid-rapidity yields ($\text{d}N/\text{d}y$) of $\pi^{\pm}$, K$^{\pm}$, $\rm K_S^0$, p($\pbar$), $\phi$, $\Lambda$($\overline{\Lambda}$), $\Xi^-$($\overline{\Xi}^+$), and $\rm K^{*0}+\overline{K}^{*0}$ measured by the STAR collaboration in Au+Au collisions at $\snn$ = 7.7, 11.5, 19.6, 27, 39, and 200~GeV~\cite{STAR:2006egk,STAR:2008med,STAR:2008bgi,STAR:2010avo,STAR:2011fbd,STAR:2017sal,STAR:2019bjj,STAR:2022hbp,STAR:2022sir,STAR:2026kqj}. Where available, data from the RHIC Beam Energy Scan II (BES-II) are preferentially employed in place of the earlier Beam Energy Scan I (BES-I), as the substantially higher statistics of the BES-II dataset reduce experimental uncertainties and thereby provide tighter constraints on the extracted thermal parameters. Fits were generally performed in the 0–10\%, 10–20\%, 20–40\%, and 40–80\% centrality classes; however, at 200~GeV (0–10\%, 10–40\%, 40–60\%, 60–80\%), the centrality intervals were chosen to match those in which the data were reported. When the experimental $\text{d}N/\text{d}y$ values were available only in finer centrality bins, they were combined using a centrality-bin-width weighted average to obtain the yields corresponding to the desired intervals. 
In Au+Au collisions at $\snn=200$ GeV, the STAR experiment has not published 5–10\% centrality yields for $\rm K_S^0$, $\Lambda$($\overline{\Lambda}$), and $\Xi^-$($\overline{\Xi}^+)$, which are needed to reconstruct the 0–10\% $\text{d}N/\text{d}y$ yields used in our analysis. We resolve this problem by fitting the measured yields as a function of average number of participants ($\langle N_{\mathrm{part}} \rangle$) using a power-law form, $\text{d}N/\text{d}y = a\langle N_{\mathrm{part}} \rangle^{b}$, and evaluating the fit at the corresponding $\langle N_{\mathrm{part}} \rangle$ for the 0–10\% centrality bin. 
The associated uncertainties on the interpolated $\text{d}N/\text{d}y$ were estimated using bootstrap resampling of the fit data. Furthermore, since weak decay feed-down corrected $\text{d}N/\text{d}y$ data for $\pbar$ were not available in Au+Au collisions at $\snn = 200$~GeV, the inclusive $\text{d}N/\text{d}y$ of p$(\pbar)$ were used in the analysis. The \texttt{Thermal-FIST} package allows for explicit control over individual decay channels for each hadron species; accordingly, the model was configured to include weak decay contributions in addition to the primordial (anti-)protons during the fitting procedure. For brevity, we denote $\rm K^{*0}+\overline{K}^{*0}$ as $\kstar$, and particle symbols refer to both particles and antiparticles unless otherwise stated.

In the present study, we consider resonances with decay widths $\Gamma > 15$~MeV (corresponding to lifetimes $\tau \lesssim 13.2$~fm/$c$) to be short-lived, whereas all other hadrons, which include both stable hadrons and long-lived resonances such as $\phi$, are treated as chemically frozen and do not undergo regeneration or re-scattering in the hadronic phase. The hadronic fireball is modeled as an ideal gas of hadrons and resonances in the grand canonical ensemble, without incorporating excluded-volume or other interaction corrections. We use PDG2020~\cite{ParticleDataGroup:2020ssz} as the input hadronic spectrum. Thermal fits are performed using five free parameters: $\Tch$, $\mu_B$, R, $\Tkin$, and strangeness suppression factor ($\gamma_S$). The electric charge and strangeness chemical potentials, $\mu_Q$ and $\mu_S$, are determined by imposing global conservation constraints corresponding to a net charge-to-baryon ratio of $Q/B = 0.4$ and vanishing net strangeness ($S = 0$). The electric charge fugacity factor, $\gamma_Q$, is fixed to unity.

Calculations are carried out within the conventional zero-width approximation, wherein all resonances are treated as stable particles with fixed pole masses. However, the effects of finite resonance widths may have a sizable influence on fit quality~\cite{Vovchenko:2018fmh} and, potentially, the values of extracted parameters. Therefore, we have also examined the effects of incorporating finite resonance widths, with the corresponding results summarized in Appendix~\ref{appendix:finitewidth}.

\subsection{Baryon-Antibaryon Annihilation and \texorpdfstring{$\bm{\bar{\text{p}}}$/p}{pbar/p} Suppression}

Beyond resonance suppression, the HRG-PCE framework can also accommodate residual inelastic interactions that persist below $\Tch$. In particular, the $B\overline{B} \leftrightarrow n\pi$ annihilation reaction continues to play a crucial role by depleting (anti-)baryon yields during the hadronic phase. The \texttt{Thermal-FIST} framework includes nucleon-antinucleon annihilation processes, $\text{N}\overline{\text{N}} \leftrightarrow \langle n_\pi^{\text{N}\overline{\text{N}}} \rangle\, \pi$, where N represents p or n, and $\pi$ includes all three pion charge states. We adopt $\langle n_\pi^{\text{N}\overline{\text{N}}} \rangle = 5$ based on Ref.~\cite{Dover:1992vj}.

In this scenario, the total yields of N, $\overline{\text{N}}$, and $\pi$ are no longer conserved individually. Instead, the evolution of the system conserves two effective quantities~\cite{Vovchenko:2022xil}:
\begin{align}
    N^{\text{tot}}_{\text{ann}} &= \frac{N_\text{N}^{\text{tot}} + N_{\overline{\text{N}}}^{\text{tot}}}{2} + \frac{N_{\pi^+} + N_{\pi^-} + N_{\pi^0}}{\langle n_\pi^{\text{N}\overline{\text{N}}} \rangle}, 
    \label{eq:BBbar1}\\
    N_\text{N}^{\text{net,tot}} &= N_\text{N}^{\text{tot}} - N_{\overline{\text{N}}}^{\text{tot}}.
    \label{eq:BBbar2}
\end{align}
Assuming that annihilation and regeneration reactions proceed in relative chemical equilibrium, the following relation among chemical potentials holds:
\begin{equation}
    \mu_\text{N} + \mu_{\overline{\text{N}}} = \langle n_\pi^{\text{N}\overline{\text{N}}} \rangle \overline{\mu}_\pi, 
    \qquad \text{with} \quad \overline{\mu}_\pi = \frac{\mu_{\pi^+} + \mu_{\pi^-} + \mu_{\pi^0}}{3},
    \label{eq:PCE2}
\end{equation}
with $\mu_\text{N} = \mu_\text{p}$ and $\mu_\text{N} = \mu_\text{n}$ treated separately. In addition to the annihilation-related constraints, conservation of electric charge and isospin is maintained in the \texttt{Thermal-FIST} by fixing the corresponding net pion number differences during the hadronic evolution,
\begin{align}
    N_{\pi^+} - N_{\pi^-} &= \text{const}, \label{eq:chargeCons} \\
    N_{\pi^+} - N_{\pi^0} &= \text{const}, \label{eq:isospinCons}
\end{align}
and likewise the net number of final protons,
\begin{align}
\label{eq:Npcons}
    N_{\rm p} - N_{\pbar} &= \text{const}, 
\end{align}
thereby ensuring that the net charge and isospin asymmetries established at chemical freeze-out are preserved during the hadronic evolution.

Together, the constraints in Eqs.~(\ref{eq:BBbar1}--\ref{eq:isospinCons}) allow for the determination of the chemical potentials $\mu_\text{N}$, $\mu_{\overline{\text{N}}}$ (resolved into $\mu_\text{p}$, $\mu_{\bar{\text{p}}}$, $\mu_\text{n}$, $\mu_{\bar{\text{n}}}$), and $\mu_\pi$ (resolved into $\mu_{\pi^+}$, $\mu_{\pi^-}$, and $\mu_{\pi^0}$) at a given temperature $\text{T} < \Tch$ in the hadronic phase.

Baryon-antibaryon annihilation can have a significant effect on the yields of (anti)protons and pions. In Ref.~\cite{Vovchenko:2022xil}, the centrality dependence of the p/$\pi$ ratio was used to estimate the temperature $\Tann$ for the freeze-out of baryon annihilation in Pb+Pb collisions at $\snn$ = 5.02~TeV. It was found that $\Tann$ lies between the kinetic and chemical freeze-out, $\Tkin < \Tann < \Tch$, i.e. baryon–antibaryon annihilation reactions cease before the end of the hadronic phase. In baryon-rich matter probed at RHIC energies, baryon annihilation has a strong impact on the ratio of antibaryons to baryons, which provides a cleaner probe of this reaction compared to the p/$\pi$ ratio. For this reason, in this work, we study the centrality dependence of the $\pbar$/p ratio across various collision energies to quantify the impact of baryon annihilation.

\begin{figure}
    \centering
    \includegraphics[width=1.0\linewidth]{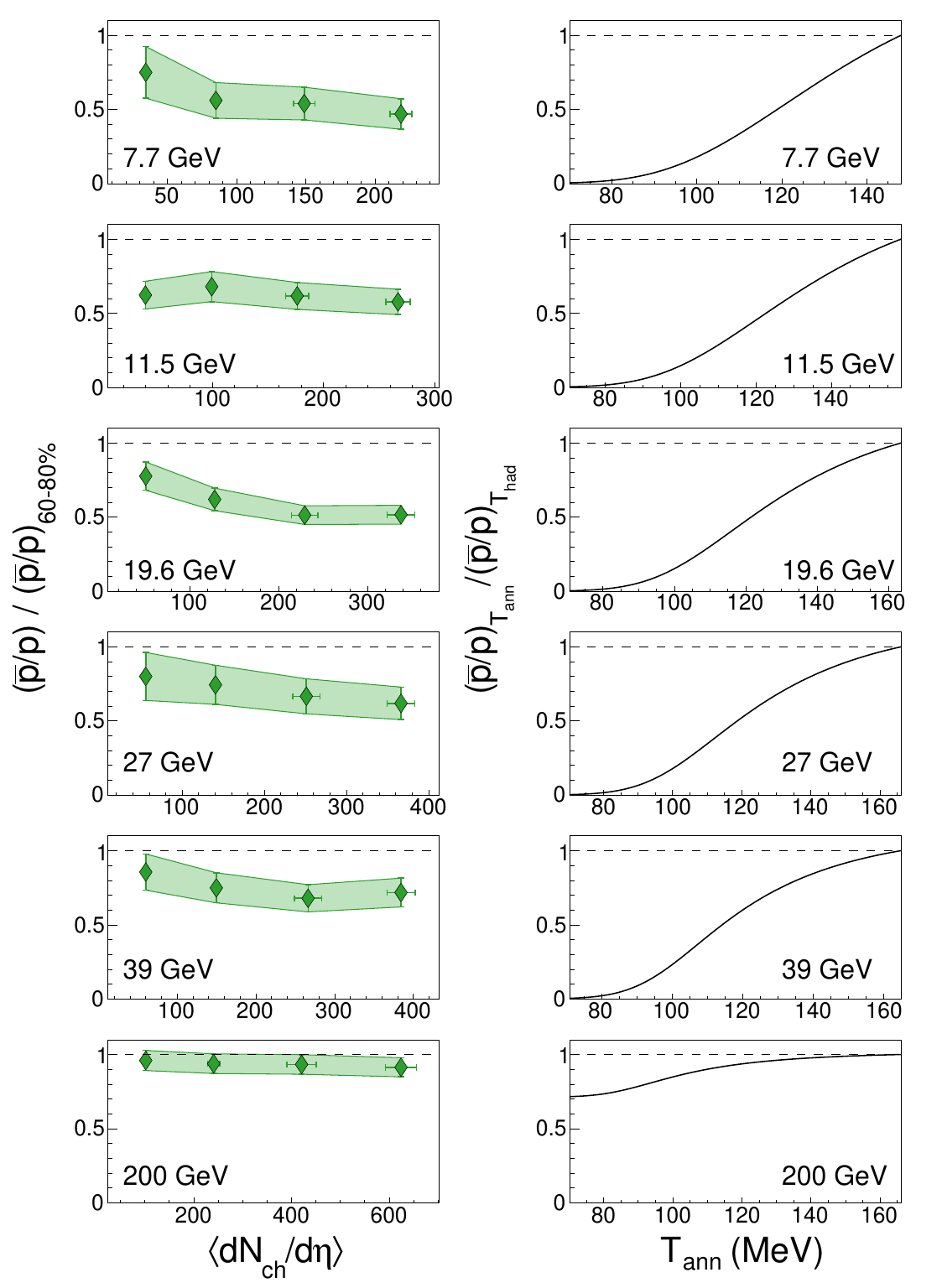}
    \caption{
    (\textit{Left}) Centrality dependence of $\pbar$/p ratio normalized to the 60–80\% peripheral value in Au+Au collisions at $\snn$ = 7.7--200~GeV. (\textit{Right}) Corresponding HRG-PCE predictions of $\pann$ as a function of temperature. Experimental values from the left panel are mapped onto the theoretical curve to extract $\Tann$.
    }
    \label{fig:ProtonRatios}
\end{figure}

The left panels of Fig.~\ref{fig:ProtonRatios} show the $\pbar$/p ratio scaled to its peripheral value (60-80\%) as a function of charged particle multiplicity ($\rm \langle dN_{\text{ch}}/d\eta \rangle$) in Au+Au collisions at $\snn$ = 7.7--200~GeV~\cite{STAR:2022hbp}. $\rm \langle dN_{\text{ch}}/d\eta \rangle$ acts as a proxy for collision centrality, with larger values corresponding to more central collisions. Only uncorrelated uncertainties in the $\text{d}N/\text{d}y$ of p and $\pbar$ are propagated to the $\pbar$/p ratio~\cite{STAR:2008med,STAR:2017sal}. Despite the sizable uncertainties, a suppression of the $\pbar$/p ratio in central collisions is evident. For Au+Au collisions at $\snn$ = 200~GeV, the inclusive $\text{d}N/\text{d}y$ of p and $\pbar$ are used, consistent with the treatment described in Sec.~\ref{subsec:PCE}.

To quantify this suppression, we estimate an effective annihilation freeze-out temperature $\Tann$ at each centrality by equating the experimentally measured $\pbar$/p, normalized to its 60–80\% value, to the HRG-PCE predictions of $\pann$ at varying temperatures (right panels of Fig.~\ref{fig:ProtonRatios}). The $\pann$ curves at each energy are computed using the hadronization temperature $\rm T_{had}$ and $\mu_B$ obtained via the fit procedure described in Sec.~\ref{subsec:PCE}, assuming $\rm T_{had} = \Tch$ and employing the most-central $\Tch$ and $\mu_B$ values at each energy. As a systematic check, we recalculated the curves using the $\Tch$ and $\mu_B$ parameterization from Ref.~\cite{Vovchenko:2015idt} and find the resulting $\Tann$ to be consistent within uncertainties. For this analysis we fix $\gamma_S = 1$, as the procedure does not directly involve strange hadrons, and all other model settings follow those specified in Sec.~\ref{subsec:PCE}.

Taken together, the resonance-suppression and baryon–antibaryon annihilation analyses leverage the HRG-PCE framework to reconstruct key aspects of the hadronic-phase evolution, from chemical freeze-out at $\Tch$, through the late-stage baryon–antibaryon annihilation processes that constrain an effective annihilation temperature $\Tann$, to the pseudo-elastic interactions governing resonance regeneration and suppression, which enable the extraction of the kinetic freeze-out temperature $\Tkin$. By anchoring these observables within the HRG-PCE model, we obtain a consistent, flow-independent characterization of freeze-out across a broad range of center-of-mass energies and collision centralities.


\section{Results}
\label{sec:results}

\begin{figure*}
    \centering
    \includegraphics[width=0.7\linewidth]{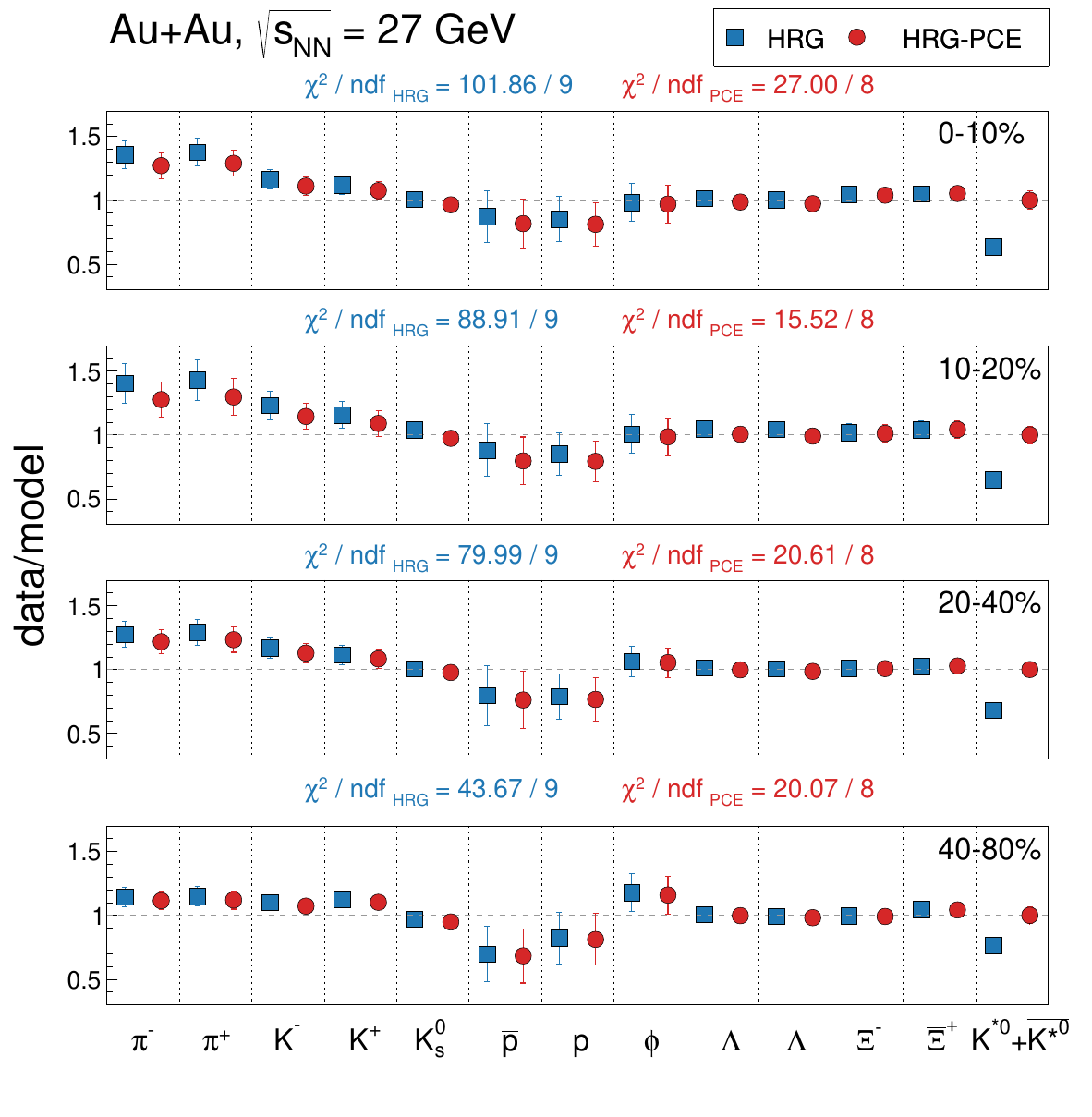}
    \caption{Ratio of experimentally measured hadron yields to thermal model fits for $\pi$, K, $\rm K_S^0$, p, $\phi$, $\Lambda$, $\Xi$, and $\kstar$ in the 0–10\%, 10–20\%, 20–40\%, and 40–80\% centrality classes of Au+Au collisions at $\snn$ = 27~GeV. Results from the standard HRG model are shown as blue square markers, while those from the HRG-PCE model are shown as red circle markers.}
    \label{fig:YieldComparison}
\end{figure*}

Figure~\ref{fig:YieldComparison} shows the ratios of experimentally measured $\text{d}N/\text{d}y$ values to the corresponding thermal model fits from HRG and HRG-PCE, for the 0–10\%, 10–20\%, 20–40\%, and 40–80\% centrality classes in Au+Au collisions at $\snn$ = 27~GeV. 
Incorporating PCE into the HRG model slightly improves agreement with stable hadron yields, particularly for pions.
The observed change in stable hadron yields under PCE does not originate from the PCE framework itself—which, by construction, preserves stable hadron yields—but rather from the modified freeze-out parameters obtained in the fitting under the PCE assumption. The improved fit quality in the HRG-PCE approach suggests that forcing short-lived resonances into a fully equilibrated HRG framework can distort the extracted freeze-out parameters. This occurs because the equilibrium framework must find a numerical compromise between the suppressed resonance yields and the thermal expectations of the stable species.

In contrast, the $\kstar$ yield is consistently better described by the HRG-PCE model, whereas the standard HRG model overestimates its abundance. This overestimation is reduced in peripheral collisions, where re-scattering effects are expected to be weaker. Similar trends are observed at other collision energies.

\begin{figure*}
    \centering
    \includegraphics[width=1.0\linewidth]{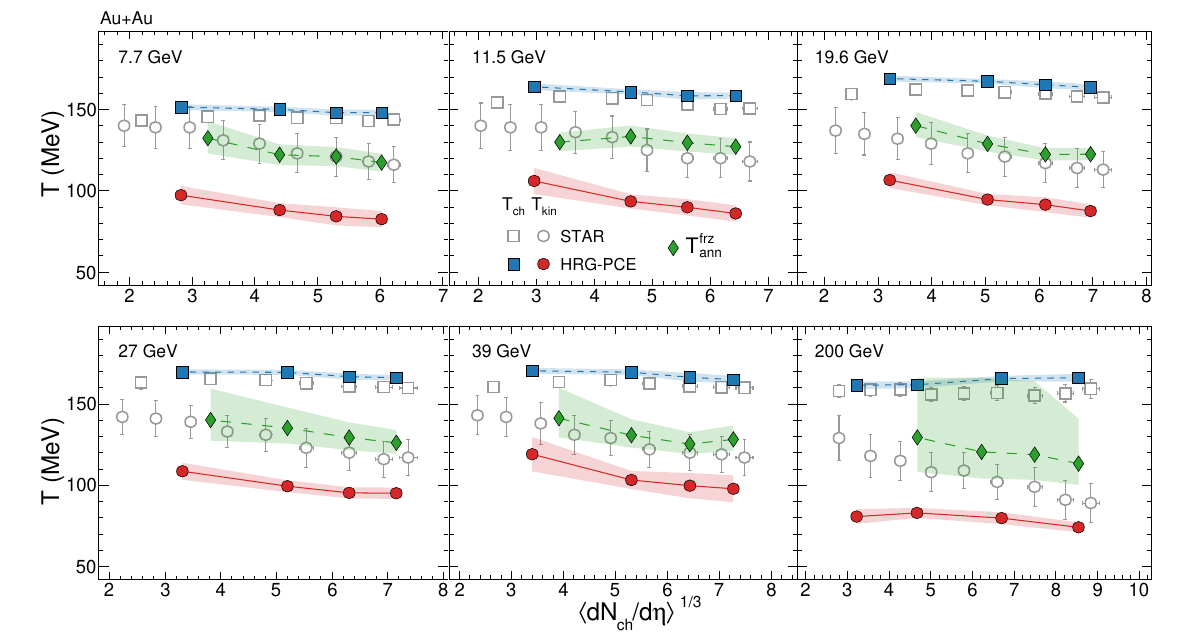}
    \caption{Chemical (solid square markers), kinetic (solid circle markers), and baryon-antibaryon annihilation (solid diamond markers) freeze-out temperatures extracted using the HRG-PCE framework as a function of charged-particle multiplicity ($\Nch$) in Au+Au collisions at $\snn$ = 7.7--200~GeV. For comparison, chemical (open square markers) and kinetic (open circle markers) freeze-out temperatures reported by STAR using \texttt{THERMUS} and blast-wave fits, respectively, are also shown~\cite{STAR:2017sal,STAR:2008med}.}
    \label{fig:TemperatureCent}
\end{figure*}

The obtained $\Tch$, $\Tkin$, and $\Tann$ freeze-out temperatures as a function of $\Nch$ in Au+Au collisions at $\snn$ = 7.7–200~GeV are shown in Fig.~\ref{fig:TemperatureCent}. For comparison, we also include $\Tch$ and $\Tkin$ values reported by the STAR collaboration from \texttt{THERMUS}~\cite{Wheaton:2004qb} and blast-wave model fits, respectively~\cite{STAR:2017sal}.

The HRG-PCE results indicate that $\Tch$ remains largely independent of centrality and is consistent with the values reported by the STAR collaboration. In contrast, $\Tkin$ shows a decreasing trend with increasing $\Nch$ and systematically falls below the values extracted from blast-wave fits.
A slight reduction in $\Tkin$ is observed in the most peripheral Au+Au collisions at $\snn$ = 200~GeV, which may be related to the corresponding decrease in the $\kstar/$K ratio reported in the most-peripheral centrality bin in Ref.~\cite{STAR:2010avo}. Furthermore, $\Tann$ shows a mild decrease with increasing collision centrality. Importantly, we find a consistent ordering, $\Tkin < \Tann < \Tch$, implying that baryon annihilation remains active during the hadronic phase but ceases prior to kinetic freeze-out.

The quantitative difference between HRG-PCE and blast-wave model fit results for $\Tkin$ is notable:
HRG-PCE results for $\Tkin$ lie consistently below the values obtained from the blast-wave model analysis by STAR~\cite{STAR:2017sal}.
We note that the blast-wave model fits rely on the shape of the $\pt$ distributions and, with few exceptions~\cite{Mazeliauskas:2018irt,Mazeliauskas:2019ifr,Melo:2019mpn}, neglect modifications of the spectra due to resonance decays and treat overall yields as free normalization parameters. 
The HRG-PCE model is agnostic to the assumptions about the flow velocity profile and the description of $\pt$ spectra, but it does rely on resonance formation and decay reactions remaining in relative chemical equilibrium until $\Tkin$, or, in other words, it equates the kinetic freeze-out stage of the bulk system with the chemical freeze-out stage of short-lived resonances. 
As both the blast-wave and HRG-PCE have distinct advantages and disadvantages, we suggest treating the difference between the two shown in Fig.~\ref{fig:TemperatureCent} as an estimate for the overall systematic error for determining $\Tkin$ in heavy-ion collisions.
We do also note that a separate blast-wave model analysis of Ref.~\cite{Melo:2019mpn}, which does take into account resonance decay effect on the shapes of the $\pt$ spectra, yields the values of $\Tkin$ which are comparable to our HRG-PCE fit results.

\begin{figure*}
    \centering
    \includegraphics[width=1.0\linewidth]{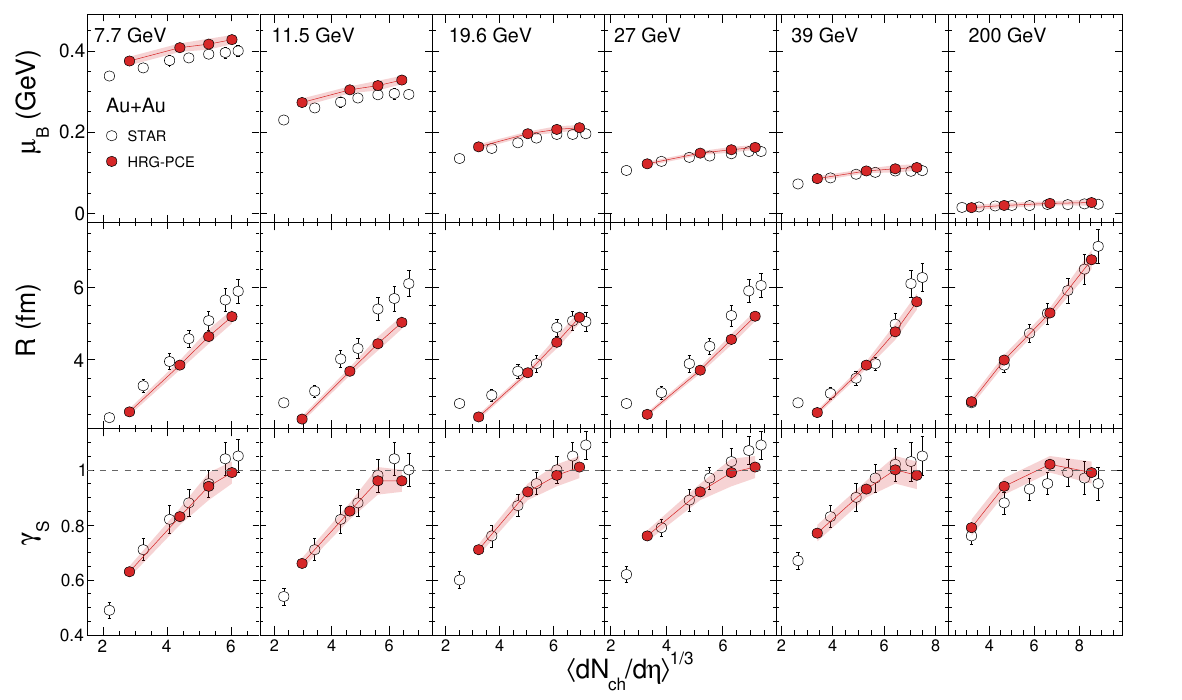}
    \caption{Baryon chemical potential, radius, and strangeness suppression factor at chemical freeze-out, extracted using the HRG-PCE framework (solid circles), as a function of charged-particle multiplicity ($\Nch$) in Au+Au collisions at $\snn$ = 7.7, 11.5, 19.6, 27, 39, and 200 GeV. For comparison, STAR values using \texttt{THERMUS} are shown as open circles~\cite{STAR:2017sal,STAR:2008med}.}
    \label{fig:CFOParCent}
\end{figure*}

Figure~\ref{fig:CFOParCent} presents the system radius R, baryon chemical potential $\mu_B$, and strangeness suppression factor $\gamma_S$ at chemical freeze-out, extracted using the HRG-PCE model, as functions of $\Nch$. Both $\mu_B$ and R increase toward more central collisions, while $\mu_B$ decreases with increasing collision energy, reflecting the reduced baryon stopping at higher $\snn$. The deviation of $\gamma_S$ from unity becomes more pronounced in peripheral collisions, reflecting incomplete chemical equilibration of strange hadrons in smaller, shorter-lived systems. Overall, the observed trends are in good agreement with those reported by the STAR collaboration using the standard HRG framework, shown as open markers.

\begin{figure*}
    \centering
    \includegraphics[width=0.9\linewidth]{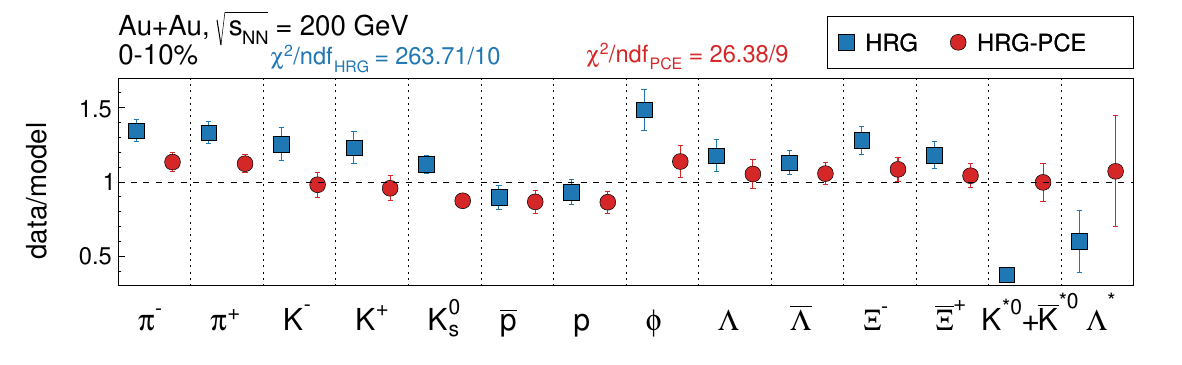}
    \caption{Ratio of experimentally measured hadron yields to thermal model fits for $\pi$, K, $\rm K_S^0$, p, $\phi$, $\Lambda$, $\Xi$, $\kstar$, and $\Lambda^*$ in 0–10\% centrality of Au+Au collisions at $\snn$ = 200~GeV. Results from the HRG and HRG-PCE models are shown as blue square and red circle markers, respectively.}
    \label{fig:200GeVallReso}
\end{figure*}

\begin{figure*}
    \centering
    \includegraphics[width=1.0\linewidth]{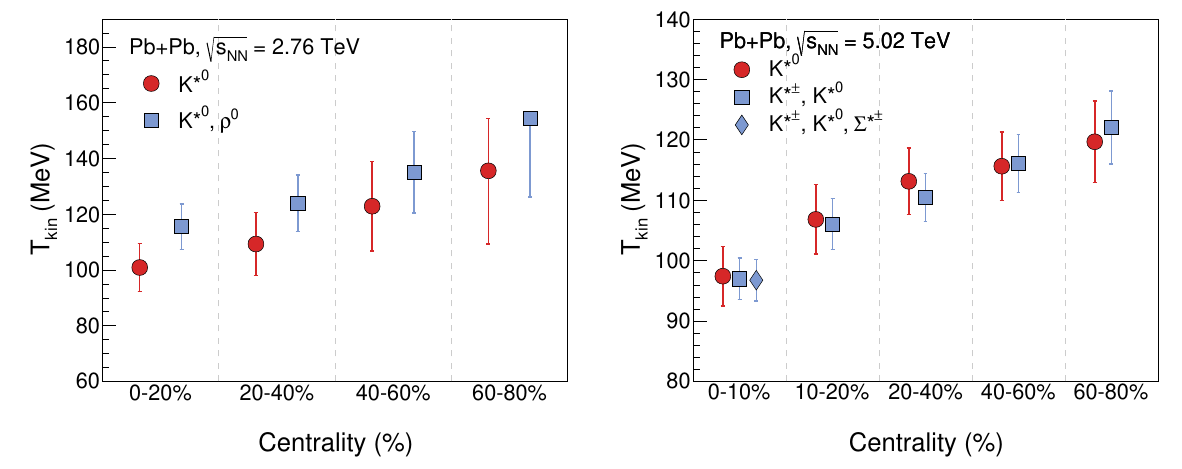}
    \caption{
    Comparison of the kinetic freeze-out temperature ($\Tkin$) extracted for various centrality classes in Pb+Pb collisions at $\snn = 2.76$~TeV (left) and $\snn = 5.02$~TeV (right) within the HRG-PCE framework. The temperatures are obtained from simultaneous fits to stable hadron yields together with different sets of short-lived resonances. For the 2.76~TeV data, $\Tkin$ is determined first using only the $\rm \kstar$ (circle markers), and then by including both $\kstar$ and $\rho^0$ (square markers). For the 5.02~TeV data, $\Tkin$ is first extracted using $\kstar$ (circle markers), followed by fits that include both $\kstar$ and $\rm K^{*\pm}$ (square markers). Finally, the effect of including $\Sigma^{*\pm}$ in the thermal fits is shown (diamond markers).}
    \label{fig:ALICE}
\end{figure*}

\begin{figure}
    \centering
    \includegraphics[width=0.99\linewidth]{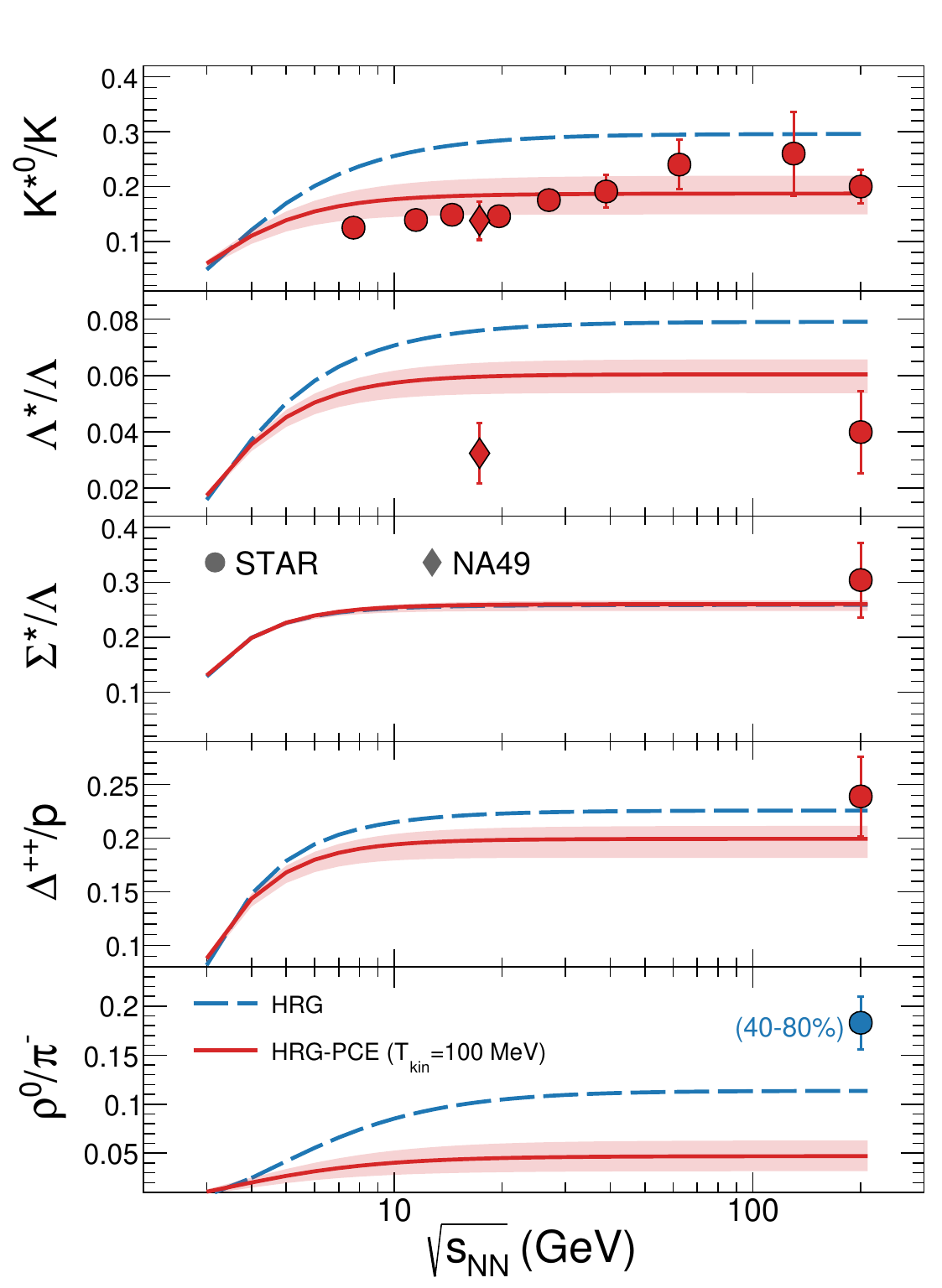}
    \caption{Predicted resonance-to-stable hadron yield ratios, $\kstar/K$, $\Lambda^*/\Lambda$, $\Sigma^*/\Lambda$, $\Delta^{++}/p$, and $\rho^0/\pi^-$, as a function of collision energy, calculated using the standard HRG model (dashed lines) and the HRG-PCE framework (solid lines) with $T_{\mathrm{kin}} = 100$ MeV. The red shaded bands represent the variation in the HRG-PCE predictions for $90~\text{MeV} < T_{\mathrm{kin}} < 110~\text{MeV}$. STAR and NA49 measurements of $\kstar/K$, $\Lambda^*/\Lambda$, and $\Sigma^*/\Lambda$ in most-central, and $\rho^0/\pi^-$ in peripheral Au+Au collisions at available energies are shown as solid markers.}
    \label{fig:ResonanceRatios}
\end{figure}

To test how the extracted kinetic freeze-out temperature depends on the choice of resonances used in the HRG-PCE fit, we incorporate $\text{d}N/\text{d}y$ of $\Lambda^*$ measured in the 0–10\% centrality class of Au+Au collisions at $\snn$ = 200~GeV~\cite{STAR:2006vhb} into the thermal fit. Figure~\ref{fig:200GeVallReso} shows the ratio of experimentally measured $\text{d}N/\text{d}y$ to the corresponding thermal model fit values from HRG and HRG-PCE. Although $\Lambda^*$ has a relatively longer lifetime ($\sim$ 13 fm/$c$), it can, within the exclusive context of the present analysis, be treated as a short-lived resonance due to our lifetime threshold of 13.2~fm/$c$. We find that the HRG-PCE model provides improved agreement with experimental data for both stable hadrons and resonances compared to the standard HRG model. The inclusion of $\Lambda^*$ has a negligible impact on the extracted kinetic freeze-out temperature.

To further test the robustness of our approach, we extend the analysis to Pb+Pb collisions at $\sqrt{s_{\rm NN}} = 2.76$ and $5.02$~TeV. For $\sqrt{s_{\rm NN}} = 2.76$~TeV, following the procedure outlined in Ref.~\cite{Motornenko:2019jha}, we fix $\mu_B = 0$~MeV and set the fugacity factors to unity, leaving $\Tch$, R, and $\Tkin$ as free parameters. Starting from a baseline fit including $\pi$, K, $\rm K_S^0$, p, $\phi$, $\Lambda$, $\Xi$, $\Omega$, and $\kstar$, we introduce $\rho^0$ into the HRG-PCE fit~\cite{ALICE:2013mez,ALICE:2017ban,ALICE:2013xmt,ALICE:2013cdo,ALICE:2018qdv}. As shown in the left panel of Fig.~\ref{fig:ALICE}, the inclusion of $\rho^0$ leads to an increase in the extracted $\Tkin$, while the values remain consistent within uncertainties. 

At $\sqrt{s_{\rm NN}} = 5.02$~TeV, we similarly fix $\Tch = 155$~MeV, $\mu_B = 0$~MeV, and fugacities to unity, as detailed in Ref.~\cite{ALICE:2023ifn}, with $\Tkin$ and R as the free parameters. The baseline fit includes stable hadrons ($\pi$, K, p, and $\phi$)~\cite{ALICE:2019hno,ALICE:2021ptz} along with $\kstar$~\cite{ALICE:2021ptz}, and is incrementally extended to include $\rm K^{*\pm}$~\cite{ALICE:2023ifn}, and finally $\Sigma^{*\pm}$ (for 0–10\% central collisions)~\cite{ALICE:2022zuc}. As shown in the right panel of Fig.~\ref{fig:ALICE}, the extracted $\Tkin$ remains consistent across all scenarios within uncertainties, suggesting that the inclusion of these additional resonances does not significantly alter the kinetic freeze-out temperatures.

It should be noted that the present analysis employs the conventional zero-width approximation, whereas the analyses at ALICE energies~\cite{Motornenko:2019jha,ALICE:2023ifn} adopt an energy-dependent Breit–Wigner mass distribution with constant branching ratios. The impact of these different resonance width treatments is examined in Appendix~\ref{appendix:finitewidth}, where we compare the thermal parameters extracted under the zero-width approximation to those obtained when finite resonance widths are included at both RHIC and LHC energies. The results remain qualitatively the same for different resonance width treatments. Incorporating the energy-dependent Breit–Wigner prescription slightly improves the fit quality and leads to modest shifts toward higher $\Tch$ and lower $\gamma_S$, while the remaining parameters are stable within uncertainties. 

Together, these results demonstrate that the thermal freeze-out parameters inferred using $\kstar$ yields offer a reliable probe of the late-stage hadronic interactions. The analysis framework remains broadly applicable and naturally accommodates additional resonance species as relevant experimental data become available. We note that the assumption of relative chemical equilibrium of resonance decays and regeneration can be relaxed by employing rate equations, as was recently done in Ref.~\cite{Neidig:2025xgr} for studying the $\kstar/$K ratio. Our extracted values of $\Tkin$ are consistent with those obtained in Ref.~\cite{Neidig:2025xgr}.

Figure~\ref{fig:ResonanceRatios} shows the predicted resonance-to-stable hadron yield ratios, $\kstar/$K, $\Lambda^*/\Lambda$, $\Sigma^*/\Lambda$, $\Delta^{++}/$p and $\rho^0/\pi^-$ as a function of collision energy. These ratios are computed using both the standard HRG model at chemical freeze-out (dashed blue line) and the HRG-PCE framework at a fixed kinetic freeze-out temperature of $T_{\mathrm{kin}} = 100$ MeV (solid red line), with chemical freeze-out parameters ($\Tch$, $\mu_B$) taken from Ref.~\cite{Vovchenko:2015idt}. The red shaded band corresponds to HRG-PCE calculations with $\Tkin$ variations between 90 and 110 MeV. We find that ratios involving short-lived resonances such as $\kstar$, $\rho^0$, and $\Delta^{++}$ exhibit sizable suppression in the HRG-PCE predictions compared to the standard HRG model due to re-scattering in the hadronic phase. In contrast, the $\Sigma^*/\Lambda$ ratio shows little to no suppression despite the short lifetime of the $\Sigma^*$ (lifetime $\sim5$ fm/$c$), likely due to efficient regeneration. These predictions are compared against available measurements from the STAR collaboration in most-central Au+Au collisions, except for the $\rho^0/\pi^-$ ratio, which has only been measured in the 40–80\% centrality class~\cite{STAR:2002npn,STAR:2003vqj,Zhang:2004rj,STAR:2006vhb,STAR:2010avo,STAR:2022sir,STAR:2026kqj}. We have also compared the predictions with the NA49 data of $\kstar/$K and $\Lambda^*/\Lambda$ in Pb+Pb collisions at $\snn$ = 17.3 GeV~\cite{NA49:2011bfu}.

The HRG-PCE model shows a good agreement with the measured $\kstar$/K and  $\Lambda^*/\Lambda$ ratios while the standard HRG calculations overestimate the data. The $\Sigma^*/\Lambda$ and $\Delta^{++}$/p ratios are well described by both the models. As expected, the $\rho^0/\pi^-$ ratio predicted for most-central collisions is significantly lower than the available data measured in the 40–80\% centrality class, reflecting the stronger suppression in a denser hadronic medium. A more complete picture is expected to emerge as additional data on resonance yields at RHIC energies become available in the future. These trends highlight the sensitivity of short-lived resonance yields to hadronic phase effects and the usefulness of the HRG-PCE framework in modeling them.


\section{Summary}
\label{sec:summary}

We performed a thermal model analysis of hadron production in Au+Au collisions across a wide range of center-of-mass energies and centralities using both the standard HRG model and the HRG-PCE framework, which incorporates chemical evolution in the hadronic phase. Our primary analysis considers two freeze-out stages: (i) the conventional chemical freeze-out stage, where the temperature $\Tch$ is mainly determined by the yields of stable hadrons, and (ii) the kinetic freeze-out stage, characterized by the freeze-out temperature $\Tkin$, which is sensitive to the yields of short-lived resonances. In addition, we also consider the freeze-out of baryon annihilation reactions at an intermediate stage, at $\Tkin < \Tann < \Tch$, which is most sensitive to $p/\pi$ ratio at LHC energies and to $\pbar$/p ratio at RHIC energies.

By fitting hadron and resonance yield data from the STAR collaboration, we extract the chemical ($\Tch$) and kinetic freeze-out ($\Tkin$) temperature, along with baryon-chemical potential ($\mu_B$), fireball radius (R), and strangeness suppression factor ($\gamma_S$) at chemical freeze-out. We find that $\Tch$ remains approximately constant with centrality and agrees with previous measurements from the STAR collaboration, while $\Tkin$ decreases with increasing centrality, and lies below values obtained from blast-wave fits, indicating continued hadronic interactions after chemical decoupling.

The HRG-PCE model, which includes pseudo-elastic processes such as resonance regeneration and suppression, significantly improves the description of short-lived resonance yields, particularly $\kstar$. Inclusion of the $\Lambda^*$ in the thermal fit in most central Au+Au collisions at $\snn$ = 200 GeV did not alter the extracted freeze-out parameters. A cross-check using Pb+Pb collisions at $\snn$ = 2.76 and 5.02~TeV shows that $\Tkin$ values remain stable within uncertainties even when additional resonances like $\rho^0$, $\rm K^{*\pm}$, and $\Sigma^{*\pm}$ are included in the fit, underscoring the robustness of the method.

Predictions of resonance-to-stable hadron yield ratios over a broad energy range demonstrate the suppression of ratios involving short-lived resonances such as $\kstar$, $\rho^0$, and $\Delta^{++}$ in the HRG-PCE model compared to the standard HRG. The $\Sigma^*/\Lambda$ ratio, despite the short lifetime of $\Sigma^*$, shows no significant suppression. The HRG-PCE model shows good agreement with the experimental data.

Beyond resonance observables, we study baryon-antibaryon annihilation processes by analyzing the suppression of the $\pbar$/p ratio in central collisions. This suppression offers an experimental handle to quantify the cessation of inelastic baryon-antibaryon reactions via an effective annihilation freeze-out temperature ($\Tann$). Assuming relative chemical equilibrium in baryon-antibaryon annihilation ($B\overline{B} \leftrightarrow n\pi$), we determine $\Tann$ by matching the measured $\pbar$/p ratios, normalized to peripheral values, to HRG-PCE predictions. The results show that $\Tann$ lies between $\Tkin$ and $\Tch$, indicating that baryon annihilation persists during the hadronic phase and ceases before kinetic freeze-out. This result is consistent with earlier findings at LHC energies~\cite{Vovchenko:2022xil}.

Overall, this study provides a unified, flow-independent reconstruction of the hadronic thermal evolution, from chemical freeze-out through baryon–antibaryon annihilation and resonance-driven pseudo-elastic processes to kinetic decoupling, anchored entirely within the HRG-PCE framework.

Building on recent developments in thermal modeling of the hadronic phase, this study demonstrates how a thermodynamic approach can characterize chemical freeze-out, baryon-antibaryon annihilation, and kinetic decoupling from hadron yield data. While $\Tann$ and $\Tkin$ are extracted independently in this analysis, a more complete treatment would evolve the system from chemical freeze-out through a phase of annihilation down to $\Tann$, followed by a standard HRG-PCE evolution to $\Tkin$. This remains an area of interest for future studies. Taken together, the results offer a consistent and data-driven thermal narrative of the late-stage hadronic evolution, laying groundwork for more dynamical modeling of heavy-ion collisions within a thermal framework.

\begin{acknowledgments}
CJ acknowledges the financial support from DAE-DST, Government of India, bearing Project No. SR/MF/PS-02/2021-IISERT (E-37130). VV was supported by the U.S. Department of Energy, Office of Science, Office of Nuclear Physics, Early Career Research Program under Award Number DE-SC0026065.
\end{acknowledgments}


\appendix
\renewcommand{\thefigure}{\thesection\arabic{figure}}
\makeatletter
\@addtoreset{figure}{section}
\makeatother

\section{Effect of Finite Resonance Widths on Thermal Parameters}
\label{appendix:finitewidth}

In Ref.~\cite{Vovchenko:2018fmh}, the effect of finite resonance widths on the thermal description of hadron yields was explored at LHC energies within different resonance-width schemes. Here, we extend this investigation to RHIC energies by testing three prescriptions for modeling resonance widths in the HRG-PCE framework~\cite{Vovchenko:2018fmh}:
\begin{enumerate}
    \item \textit{eBW}: The energy-dependent Breit--Wigner scheme, where the partial decay widths $\Gamma_{i \to j}(m)$ depend on the resonance mass $m$ according to
    \begin{equation}
        \Gamma_{i \to j}(m) = b^{\mathrm{PDG}}_{i \to j}\,\Gamma^{\mathrm{PDG}}_i
        \left[\frac{1 - (m^{\mathrm{thr}}_{i \to j}/m)^2}{1 - (m^{\mathrm{thr}}_{i \to j}/m_i)^2}\right]^{L_{i \to j} + 1/2},
    \end{equation}
    valid for $m > m^{\mathrm{thr}}_{i \to j}$ and zero otherwise, where $m^{\mathrm{thr}}_{i \to j}$ is the sum of the masses of all decay products in the channel $j$. Here $b^{\mathrm{PDG}}_{i \to j}$ and $\Gamma^{\mathrm{PDG}}_i$ denote the PDG branching ratio and total width of the resonance $i$, respectively, and $L_{i \to j}$ is the orbital angular momentum released in the decay channel. The total width is given by
    \begin{equation}
        \Gamma_i(m) = \sum_j \Gamma_{i \to j}(m),
    \end{equation}
    and the corresponding mass-dependent branching ratios are obtained as
    \begin{equation}
        b_{i \to j}(m) = \frac{\Gamma_{i \to j}(m)}{\Gamma_i(m)}.
    \end{equation}
    This energy dependence of the branching ratios modifies the decay feed-down to stable hadrons, suppressing low-mass contributions near the decay thresholds for broad resonances.
    
    \item \textit{eBWconstBR}: The same as \textit{eBW}, except that the decay branching ratios are assumed to remain constant, independent of the resonance mass.
    
    \item \textit{ZeroWidth}: The conventional zero-width approximation, where all resonances are treated as stable particles with fixed pole mass $m_i$,
    \begin{equation}
        \rho_i(m) = \delta(m - m_i).
    \end{equation}
\end{enumerate}

The total hadron yields in each scheme are calculated as
\begin{equation}
    \langle N_i^{\text{tot}} \rangle = \langle N_i^{\text{prim}} \rangle + \sum_R \langle n_i \rangle_R \, \langle N_R^{\text{prim}} \rangle,
\end{equation}
where the feed-down term $\langle n_i \rangle_R$ reflects the resonance decay kinematics according to the chosen width prescription.

\subsection*{Observations at RHIC Energies}

\begin{figure*}
    \centering
    \includegraphics[width=0.7\linewidth]{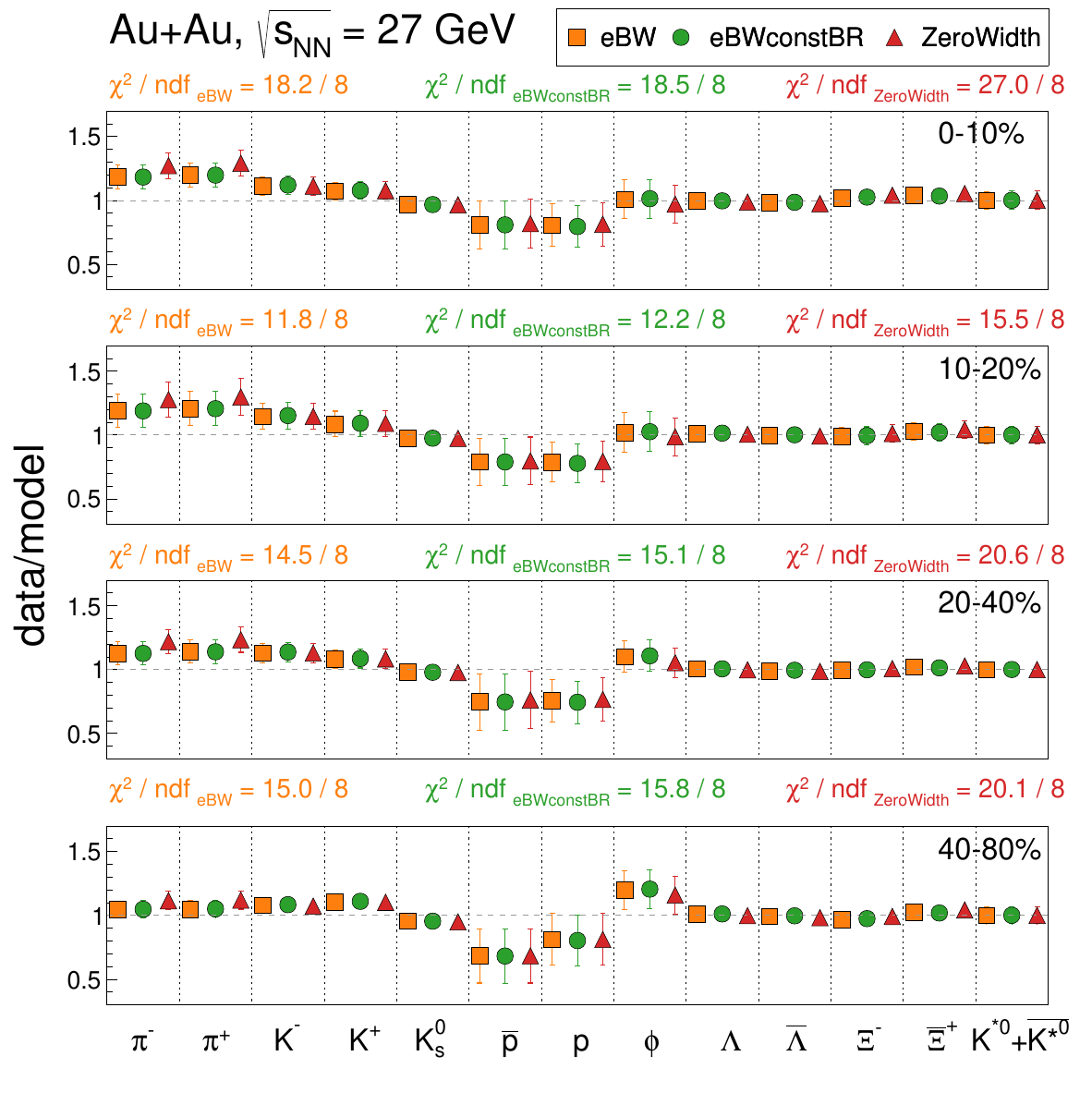}
    \caption{Data/model ratios for measured mid-rapidity yields relative to HRG-PCE calculations in Au+Au collisions at $\snn$ = 27 GeV for different centrality classes. Results are compared for three resonance-width prescriptions: the energy-dependent Breit–Wigner (\textit{eBW}), the energy-dependent Breit–Wigner with constant branching ratios (\textit{eBWconstBR}), and the zero-width approximation (\textit{ZeroWidth}).}
    \label{Afig:yield_res_width}
\end{figure*}

Figure~\ref{Afig:yield_res_width} shows the ratios of experimentally measured $\text{d}N/\text{d}y$ values to the corresponding HRG-PCE model predictions for Au+Au collisions at $\snn$ = 27~GeV, in the 0–10\%, 10–20\%, 20–40\%, and 40–80\% centrality classes, under three different treatments of resonance widths: \textit{eBW}, \textit{eBWconstBR}, and \textit{ZeroWidth}. The \textit{eBW} and \textit{eBWconstBR} schemes provide an improved description of the data, particularly for pions, resulting in smaller $\chi^2/\mathrm{ndf}$ values. A similar trend is observed across other collision energies.

\begin{figure*}
    \centering
    \includegraphics[width=1.0\linewidth]{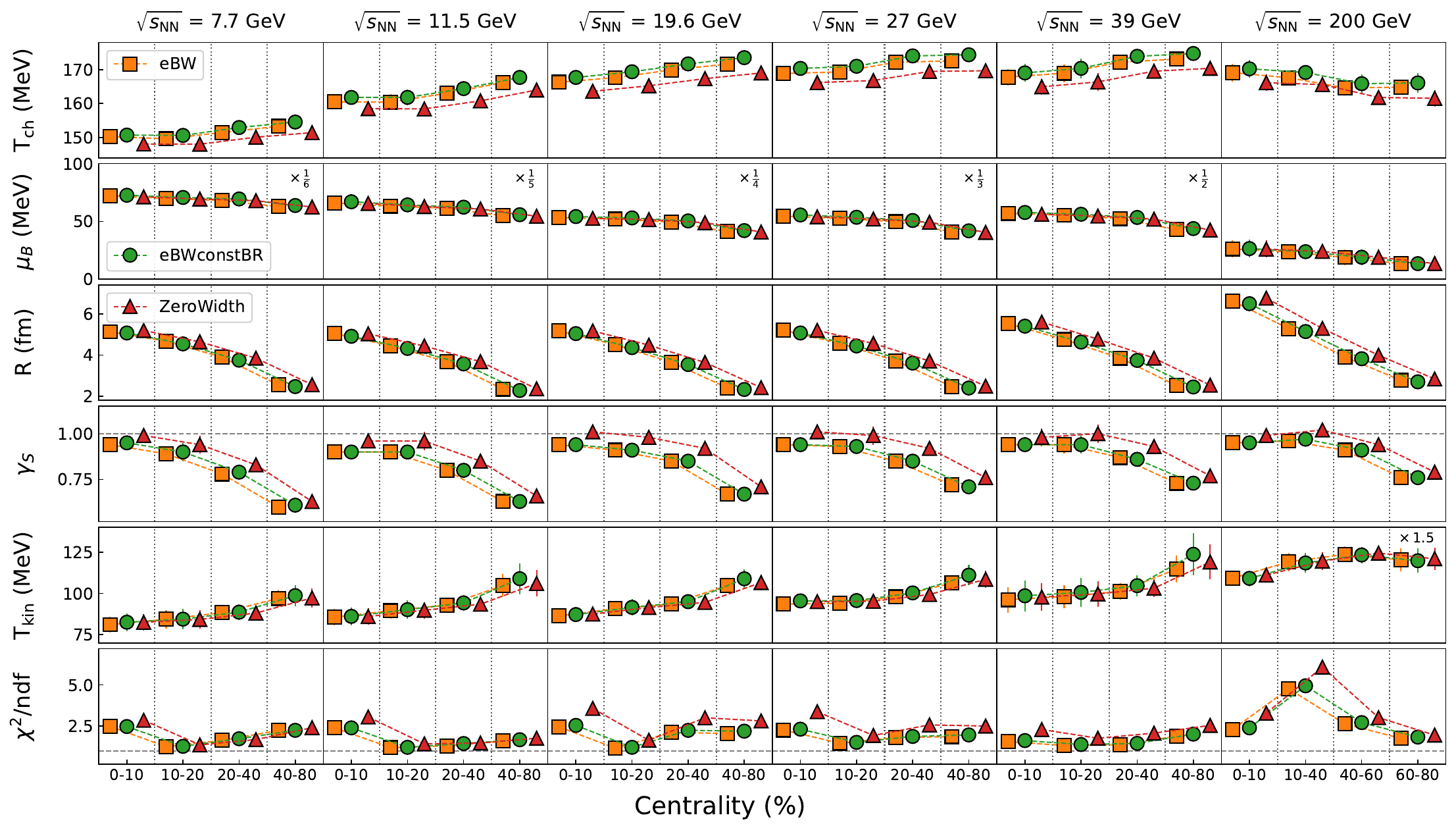}
    \caption{Centrality dependence of the freeze-out parameters in Au+Au collisions at $\snn$ = 7.7--200 GeV, extracted from the HRG-PCE fits to stable hadrons and the $\kstar$ resonance. The fits were performed using three different treatments of resonance widths: the energy-dependent Breit–Wigner (\textit{eBW}), the energy-dependent Breit–Wigner with constant branching ratios (\textit{eBWconstBR}), and the zero-width approximation (\textit{ZeroWidth}).}
    \label{fig:ParComparison_res}
\end{figure*}

Figure~\ref{fig:ParComparison_res} summarizes the extracted freeze-out parameters in Au+Au collisions at $\snn$ = 7.7--200~GeV for the three resonance-width scenarios. The following systematic trends are observed:
\begin{enumerate}
\item \textbf{Chemical freeze-out temperature ($\Tch$):} $\Tch$ is systematically lower in the \textit{ZeroWidth} case across all centralities. A slight increase of $\Tch$ towards peripheral collisions is observed, except at $\snn$ = 200~GeV, where the trend is less pronounced.
\item \textbf{Baryochemical potential ($\mu_B$):} The extracted $\mu_B$ values are consistent within uncertainties among the three schemes, indicating that $\mu_B$ is largely insensitive to the resonance-width treatment.
\item \textbf{Fireball radius (R):} The extracted radii are consistent within uncertainties across all schemes.
\item \textbf{Strangeness suppression factor ($\gamma_S$):} $\gamma_S$ is systematically higher in the \textit{ZeroWidth} approximation.
\item \textbf{Kinetic freeze-out temperature ($\Tkin$):} Within uncertainties, $T_\mathrm{kin}$ is consistent among the three scenarios and shows a gradual increase towards peripheral collisions.
\item \textbf{Fit quality ($\chi^2/\mathrm{ndf}$):} The $\chi^2/\mathrm{ndf}$ values consistently favor the \textit{eBW} and \textit{eBWconstBR} schemes, whereas the \textit{ZeroWidth} approximation yields larger $\chi^2/\mathrm{ndf}$ values, particularly at higher energies.
\end{enumerate}

\subsection*{Observations at LHC Energies}
\label{appendix:finitewidthLHC}

To further assess the sensitivity of the extracted kinetic freeze-out temperature to the treatment of resonance widths, we repeat the HRG-PCE fits for Pb+Pb collisions at $\snn = 2.76$ and $5.02$ TeV using the \textit{eBWconstBR} prescription. This extends the comparison carried out in Sec.~\ref{sec:results}, where the same fits were performed under the \textit{ZeroWidth} approximation. Apart from the choice of width treatment, all model configurations are kept identical to those described in Sec.~\ref{sec:results}.

\begin{figure*}
    \centering
    \includegraphics[width=1.0\linewidth]{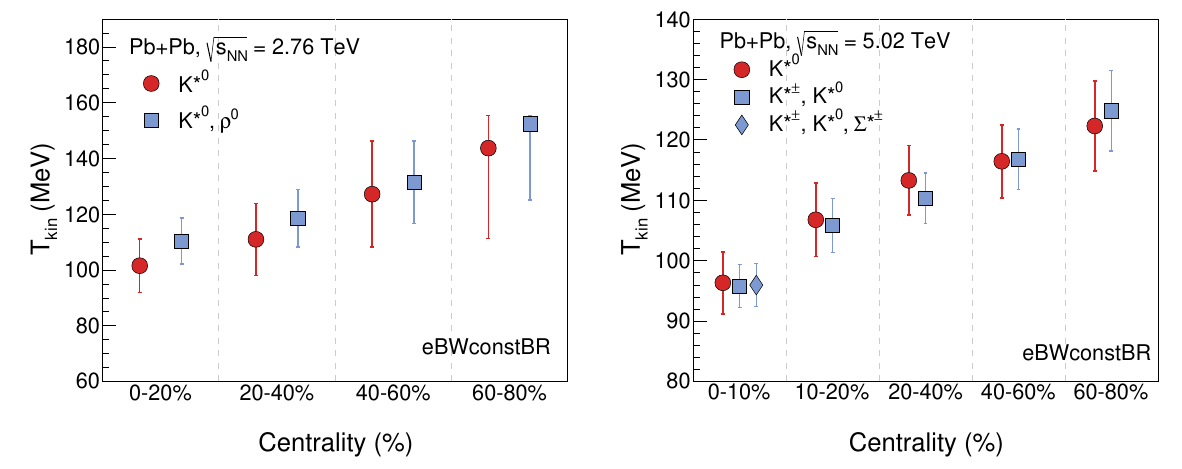}
    \caption{
    Comparison of $\Tkin$ extracted for various centrality classes in Pb+Pb collisions at $\snn = 2.76$~TeV (left) and $\snn$ = 5.02~TeV (right) within the HRG-PCE framework using the \textit{eBWconstBR} resonance-width prescription. For the 2.76~TeV data, $\Tkin$ is determined first using only the $\kstar$ (circle markers), and subsequently by including both $\kstar$ and $\rho^0$ (square markers), as reported in Ref.~\cite{Motornenko:2019jha}. For the 5.02 TeV data, $\Tkin$ is initially extracted using $\kstar$ (circle markers) and compared to the ALICE results~\cite{ALICE:2023ifn} (square markers), which incorporate both $\kstar$ and $\rm K^{*\pm}$. Finally, the impact of including $\Sigma^{*\pm}$ in the thermal fits is shown (diamond markers).
    }
    \label{fig:finitewidthLHC}
\end{figure*}

Figure~\ref{fig:finitewidthLHC} presents a comparison of the kinetic freeze-out temperatures extracted using the \textit{eBWconstBR} treatment. Consistent with the observations in the \textit{ZeroWidth} scenario, the $\Tkin$ values obtained when the $\rho^{0}$ meson is included in the HRG-PCE fits along with $\kstar$ are systematically higher, yet remain consistent with the baseline results within uncertainties. Furthermore, at $\snn = 5.02$~TeV, the extracted temperatures remain consistent when the resonance set used in the fit is extended to include $\rm K^{*\pm}$ and $\Sigma^{*\pm}$. This confirms that the extraction of $\Tkin$ is robust against the choice of resonance width prescription at LHC energies.

\clearpage

\bibliographystyle{apsrev4-1}
\bibliography{references}{}

\end{document}